\documentstyle[epsfig]{aipproc}

\begin{document}
\begin{flushright}
Edinburgh-PPE/97-02 \\
29th July 1997
\end{flushright}
\title{Status report of the NA48 experiment at the CERN SPS}

\author{Cinzia Talamonti$^*$ }
\address{$^*$University of Edinburgh}
\maketitle
\begin{center}
For the NA48 collaboration
\end{center}

\begin{abstract} 
The aim of the NA48 experiment at the CERN SPS is to measure direct CP violation in neutral kaon decays thus determining the parameter $\varepsilon \prime /\varepsilon$ with an accuracy of $2 \times 10^{-4}$. The advantages of NA48 with respect to previous experiments are high statistics and reduced systematic effects. The principle of the experiment and the performance of the detector components are presented. 
\end{abstract}

\section*{Introduction}
In 1964, CP violation was discovered in the decay of the long-lived kaon into two pions \cite{cpv}. Since then, considerable experimental and theoretical effort has been devoted to understanding its origin. After thirty years, no evidence of CP-violation has been observed in particle physics outside the neutral kaon system.
The cause of violation of CP symmetry is still an open issue in physics. 
The study of direct CP-violation has turned out to be difficult, both experimentally and theoretically, and the understanding of its magnitude and origin is still far from satisfactory.
The latest theoretical predictions for the value of $\Re (\varepsilon \prime /\varepsilon)$ are in the range $1\times10^{-4}\leq\Re (\varepsilon \prime /\varepsilon)\leq 15\times10^{-4}$ \cite{theo1,theo2,theo3,theo4}. The experimental measurement of $\varepsilon \prime /\varepsilon$ should allow discrimination between different theoretical hypotheses. E731 at FNAL and NA31 at CERN have measured $\Re (\varepsilon \prime /\varepsilon)$ with a precision better than $10^{-3}$. NA31 has obtained a value of $(2.0\pm0.7)10^{-3}$ \cite{na31} while E731 has found a value of $(0.74\pm 0.59)10^{-3}$ \cite{e731}. The first one indicates evidence of direct CP violation and the second is compatible with a null result.  
To clarify the situation requires new measurements, at CERN, NA48 and at FNAL, KTeV, aim for a more precise determination of $(\varepsilon \prime /\varepsilon)$ with a accuracy on $\Re (\varepsilon \prime /\varepsilon)$ of $\sim 2 \times 10^{-4}$.
\section*{Principle of the experiment }

\subsection*{The experimental technique}
Experimentally the measurement of $\varepsilon \prime /\varepsilon$ is obtained by evaluating the double ratio of the decay rates of $K_L$ and $K_S$ into two neutral and two charged pions:
\begin{eqnarray}
\Re (\varepsilon \prime /\varepsilon) \cong \frac{1}{6} \left(
1 - \frac{\Gamma(K_L \rightarrow \pi^0 \pi^0)}{\Gamma(K_S \rightarrow \pi^0 \pi^0)} \cdot \frac{\Gamma(K_S \rightarrow \pi^+ \pi^-)}{\Gamma(K_L \rightarrow \pi^+ \pi^-)}
\right)
\end{eqnarray}
with simultaneous observation of all four decay modes in the same experimental setup. The basic NA48 scheme \cite{na48} features two nearly collinear beams of $K_L$ and $K_S$, produced by protons hitting two different targets, and distinguished by tagging of the protons producing the $K_S$ component. A magnetic spectrometer with four drift chambers measures the momentum of the charged particles, and a fast liquid krypton calorimeter measures the energy, the position and the time of a neutral decay. Provided that the position of the two beams coincide in the detector, the advantage of this method is that at any given energy and vertex position, the detection efficiencies cancel. Differences and variations in detection efficiencies for $K_L$ and $K_S$ decays become unimportant, as do rate dependent effects introduced by accidental activity in the detector elements.

\subsection*{The statistical and systematic errors}
The statistical error on the double ratio R is dominated by the statistical precision of the rarest decay. In order to determine $\Re (\varepsilon \prime /\varepsilon)$ with a precision of $10^{-4}$, it is necessary to collect about $5 \times 10^{6}$ $K_L \rightarrow \pi^0 \pi^0$ decays, one order of magnitude more than the previous experiments, NA31 and E731. 

It is mostly the systematic error which limits the precision attainable on the measurement of R. In fact, due to different beam divergences, energy spectra and decay vertex positions, the acceptances of $K_L$ and $K_S$ decays are not exactly the same. In order to have a negligible difference in momentum spectra, the production angles of the $K_L$ and $K_S$ beams have been carefully chosen (2.4 and 4.2 mrad respectively). This way, the decay spectra are similar when the kaon momentum is between 70 and 170~GeV. 

Since the distribution of decay vertex are different (one is flat ($K_L$) and the other is exponential ($K_S$)), a weight W, depending on kaon momentum p and on the longitudinal coordinate z of kaon decay vertex, will be applied to $K_L$ events. Using this technique, the spatial distribution of $K_L$ decay vertices reproduces that of the  $K_S$ beam.

Another effect which can contribute to the systematic error is due to the imperfect knowledge of the fiducial volume. For the neutral decays, an accurate knowledge of the energy scale is required and this is achieved by adjusting the reconstructed longitudinal decay vertex position to a counter edge placed at the beginning of the $K_S$ beam.

Background events, mainly three-body decays, are another important source of systematic errors. In case of neutral events, most of the decays 
$K_L \rightarrow \pi^0 \pi^0 \pi^0$ are rejected by using anti-counter rings which veto photons outside the fiducial region, and an electromagnetic calorimeter with excellent resolution in energy and space. The background of charged events is given by   
$K_L \rightarrow \pi^{\pm} e^{\mp} \nu(\overline{\nu})$, 
$K_L \rightarrow \pi^{\pm}\mu^{\mp} \nu(\overline{\nu})$ and by 
$K_L \rightarrow \pi^+ \pi^- \pi^0$. The information of the muon veto counters, located at the rear end of the detector, is used at the trigger level to reject 
$K_L \rightarrow \pi^{\pm} \mu^{\mp} \nu(\overline{\nu})$ events. The invariant mass is reconstructed at the trigger level as well, in order to discard 
$K_L \rightarrow \pi^+ \pi^- \pi^0$ and 
$K_L \rightarrow \pi^{\pm} e^{\mp} \nu(\overline{\nu})$ decays (at $90\%$ level).
 
\section*{The $K_L$ and $K_S$ beams}
The NA48 experiment compares the $\pi^0\pi^0$ and $\pi^+\pi^-$ decay rates from distinct $K_L$ and $K_S$ beams which enter a common decay region along a path entirely contained in vacuum. The four decay modes are thus recorded at the same time and from the same fiducial length. The two beams are nearly collinear, converging at an angle comparable with the beam divergence. 

\begin{figure}[htp] 
\rotatebox{270}{\mbox{\epsfxsize=0.7\textwidth
\epsffile{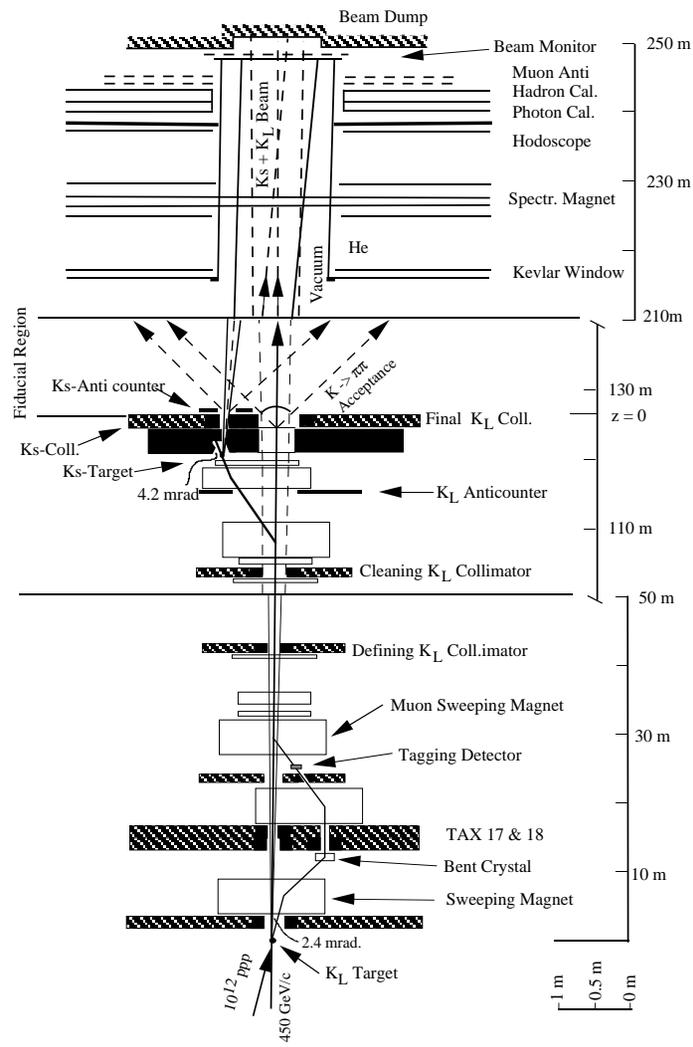}}}
\caption{The Na48 kaon beams and detector layout.}
\label{fig1}
\end{figure}

A 450~GeV/c proton beam with a nominal flux of $1.5 \times 10^{12}$ ppp strikes a beryllium target, producing the $K_L$ beam at an angle of 2.4~mrad. After a first collimator which limits acceptance, charged particles (including the remaining primary protons) are deviated from the $K_L$ line by a sweeping magnet. A fraction of the primary protons ($3 \times 10^7$ ppp) is channelled back towards the $K_L$ beam by a bent silicon crystal \cite{cry}. It then passes through a tagging counter, and is eventually deflected and focused on the $K_L$ line. It then hits a second target, placed 72~mm above the first one and at a distance of 120~m from it, to produce a $K_S$ beam at an angle of 4.2~mrad. This beam is defined by a collimator and converges with the $K_L$ one at an angle of 0.6~mrad. The exit of the $K_S$ collimator coincides with the last of three collimators designed to define the $K_L$ beam, so that background from the collimators themselves cannot reach the detector. After the $K_S$ collimator, a detector, vetoes $K_S$ decays which occurred before the collimator exit and defines the fiducial region.
A schematic view of the the kaon beam lines and of the detector layout is shown in Fig.\ref{fig1}.

\section*{The NA48 detector}

\subsection*{The tagging counter}
The $K_L$ and $K_S$ assignment for each decay is done by measuring the time difference between the passage of a proton in the tagging counter \cite{tag} upstream of the 
$K_S$ target and the event time in the detector. Events with a time difference inside of a given interval $\Delta t$ will be called ``$K_S$'', any other events will be called ``$K_L$''. It has been designed to cope with proton rates above 10MHz and to provide a detection efficiency close to 100\% combined with a time resolution better than 500~ps. To ensure that the rate in each counter is less than 1MHz the tagger is designed with two sets of staggered scintillation foils arranged alternately in the horizontal and vertical planes. The depth of the foils in the beam direction is 4~mm and the width varies from 200~$\mu$m on the beam-axis to 3000~$\mu$m at the beam edges. Thus the entire beam profile is covered. The time resolution is less than 200~ps, with a double pulse resolution below 4~ns. Inefficiency and accidental activity of the tagging counter result in a correction to R, but being decay mode independent they cannot generate a fake non-zero value for $\Re (\varepsilon \prime /\varepsilon)$, nevertheless the efficiency will be monitored at the $10^{-4}$ level.

\subsection*{The magnetic spectrometer}
A magnetic spectrometer \cite{spect} is used to analyse the $\pi^+\pi^-$ decay:
two set of drift chambers on each side of a central dipole magnet provide a measurement of charged particle momenta. Each chamber has four views X, Y, U and V to avoid ambiguities in the position measurement. Each view consist of two planes of 256 wires, at a distance of 1~cm from each other. The spatial resolution of each view is $\simeq$100$\mu$m, which corresponds to $\Delta p/p$ = 0.6\% accuracy on mean momentum measurement. The field integral of the main field component ($B_y$) is equivalent to a transverse momentum kick of 250 MeV/c. 

During the 1995 and 1996 test runs, very good performance was obtained with the magnetic spectrometer for the $K_{S,L}\rightarrow \pi^+\pi^-$ decays. The four drift chambers worked reliably and provided high detection efficiency and position resolution at rates close to 1MHz. The kaon invariant mass resolution obtained off-line from fully reconstructed $\pi^+\pi^-$ events is 3 MeV/$c^2$ in both  $K_L$ and $K_S$ beams (Fig. \ref{fig2}). The good resolution on the decay vertex position, in the transverse plane, provides a clear separation of the $K_L$ and $K_S$ sources, allowing a measurement of the tagging efficiency in the charged decay mode.

\subsection*{The charged hodoscope}
The timining information provided by this detector is used in conjunction with the tagger to identify the $K_L$ or $K_S$ origin of charged decays, and also in the trigger to select two-body decays. The detector consist of two planes of scintillators counters, at a distance of 50~cm, located after the magnetic spectrometer along the beam line. Each plane consists of 64 scintillators, arranged vertically and horizontally.
The time resolution for a two pion event is better than 250~ps.

\subsection*{The electromagnetic calorimeter}
\begin{center}
\begin{figure}[t] 
\mbox{\epsfxsize=.45\textwidth\epsffile{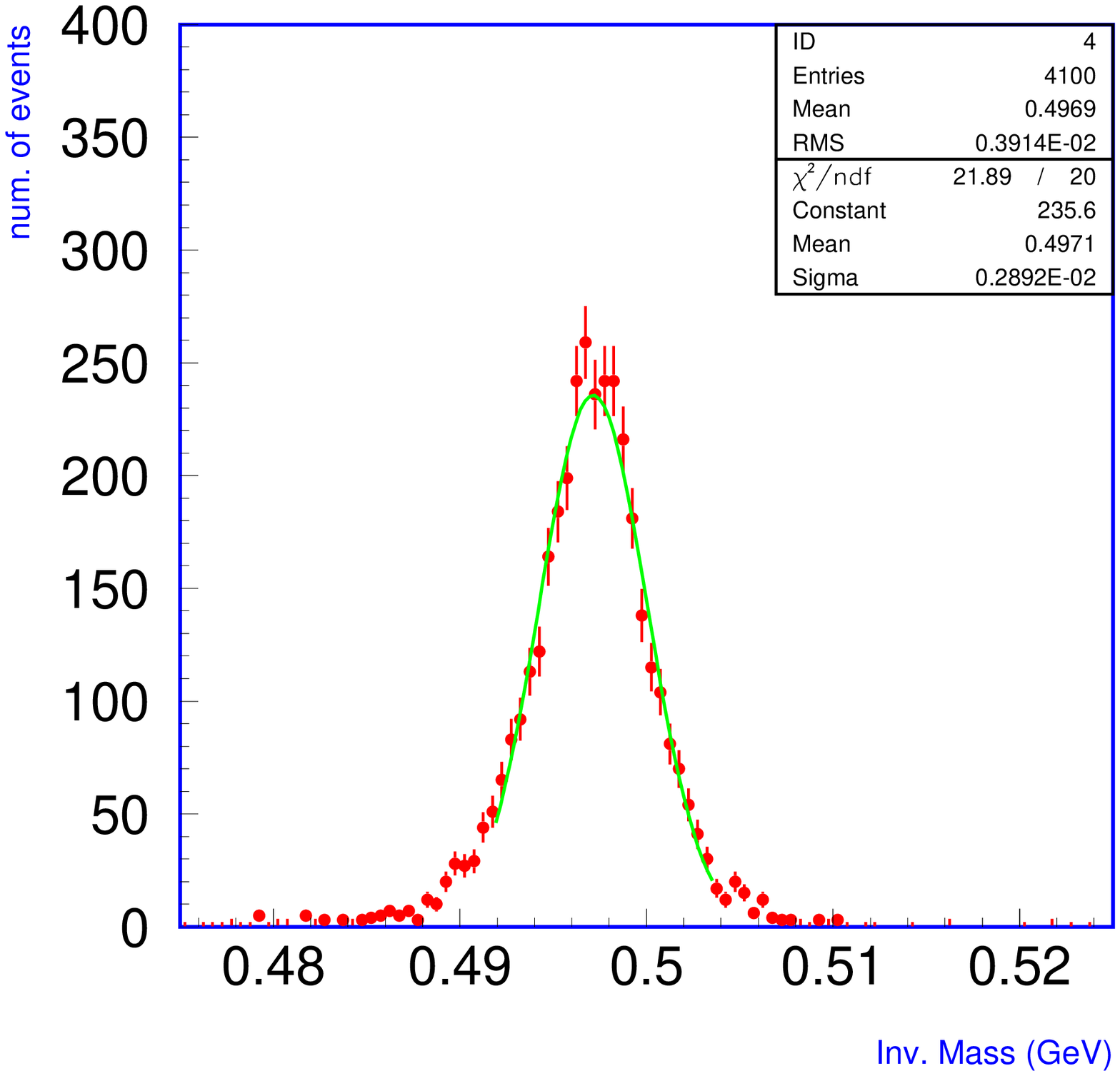}}
\hfill
\mbox{\epsfxsize=.45\textwidth\epsffile{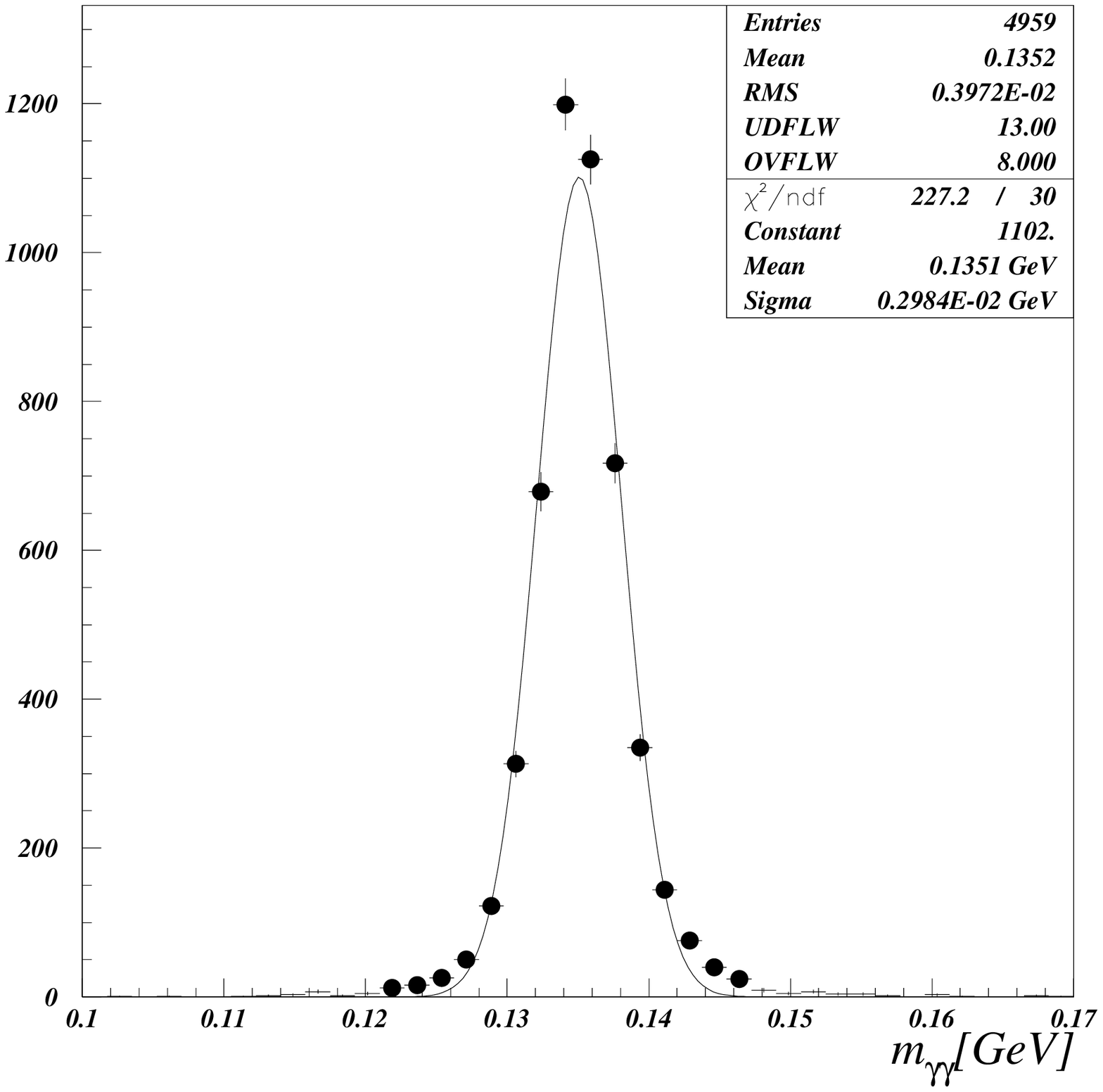}}
\vspace*{0.5 cm}
\caption{Invariant mass distributions of the $\pi^+ \pi^-$ measured in the magnetic spectrometer and of two photons measured in the calorimeter.}
\label{fig2}
\end{figure}
\end{center}

The characteristic of the decays to be detected, and the high rate of the experiment, impose same stringent requirements on the electromagnetic calorimeter. In addition, a large sensitive area ($\sim 6 m^2$) has to be covered. Good energy resolution is needed: $\sigma_E/E \sim 1\%$ at 10~GeV with a constant term lower than 0.5\%. A spatial resolution $\sigma_{x,y} \sim 1.5~mm$ is also needed.
The tranverse scale has to be accurate to 0.1/1000~mm to keep the systematic error induced by a possible difference in energy scale for charged and neutral decays at the desired level. For $K_S$ identification a timing accurancy $\sigma_t \leq 0.5~ns$ has to be achieved. High single rate capability($\sim 1 MHz$) is needed to collect the required statitiscs, high granularity and short sensitive time would reduce the effect of accidentals and background. The NA48 collaboration has built a fully sensitive liquid krypton ionization chamber with electrodes parallel to the shower direction and a tower structure segmentation into $2 \times 2 cm^2$ cells. Its transverse dimension is about 2.6m and its thickness is 125cm, corresponding to about 27 $X_0$ of liquid krypton. The total krypton volume in the cryostat is about $10m^3$. To suppliment the time measurement of the detected photons, scintillating fibers have been inserted vertically in the liquid krypton at a depth of 9.5 $X_0$.
The required performance has been achieved after several test beam runs with a prototype \cite{cpro}. The energy resolution achieved with an electron test beam is 
\begin{eqnarray}
\sigma_E / E = 3.5\% / \sqrt{E} \oplus 40 MeV/E \oplus .42\%
\end{eqnarray}              
During the 1996 test run period only the 8\% of the LKr electronics was available; the entire calorimeter was read-out by instrumenting ``supercells'' of 2x8 single cells. In this way physics data could be taken with both beams. The performance of the calorimeter was tested during the calibration period with an electron beam and the resolution obtained is in good agreement with prototype results. The neutral decays in two $\pi^0$ with four clusters in the Lkr calorimeter were identified and their rates were found to agree with the predicted ones. The $\pi^0$ mass resolution is about 3.5~MeV/$c^2$ (Fig.\ref{fig2}) due to the poor granularity of the supercells and their associated noise, but it is compatible with the proposal (1~MeV/$c^2$).

\subsection*{The hadronic calorimeter}
The most important task of this subdetector is to provide an energy threshold signal which is part of trigger, and is used to reject background coming from the three body decays. The front and the back modules consist of alternate iron and scintillator planes oriented either horizontally or vertically. The thickness corresponds to 7.2 interaction lengths. The energy resolution of the hadron calorimeter is 65\%/$\sqrt E$. 
\subsection*{The muon veto system}
To reject the background due to the $K_L \rightarrow \pi^{\pm}\mu^{\mp} \nu(\overline{\nu})$ it is used the information provided by the $\mu$-counters, located at the end of the beam line. A total of 28 scintillator strips are arranged in three planes, preceded by an iron layer 80~cm thick. The strips are alternatively vertical and horizontal, partially overlapping. The efficiency of this system is better than 99\% for muons above 5~GeV.
\section*{ Perspectives and conclusion}
The beam and detector for the NA48 experiment are ready to take data for the study of direct CP violation in the $K^0$ system. Data taking will start in August. The beams have achieved their design goals, including the novel use of a bent silicon crystal for deflecting protons at 450~Gev/c and tagging them with time resolution of better than 500~ps at 10~MHz beam rates. The detector is also complete, the required mass resolutions have been achieved in the magnetic spectrometer, and will be achieved in the liquid krypton calorimenter when the full electronics is commissioned in July 1997.

\end{document}